Stresa, Italy, 25-27 April 2007# DESIGN AND MODELING OF MICROMECHANICAL GaAs BASED HOT PLATE FOR GAS SENSORS

*J. Jakovenko [1], M. Husǂk [1], T. Lalinský [2], M. Držʼk [3], G. Vanko [2]*

[1] Dept. of Microelectronics
Czech Technical University in Prague
Technicka 2, 166 27, Prague 6,
Czech Republic,
jakovenk@feld.cvut.cz    http://www.micro.feld.cvut.cz
[2] Institute of Electrical Engineering, Slovak Academy of Sciences, Bratislava, Slovakia
[3] International Laser Center, Ilkovičova 3, 812 19 Bratislava, Slovakia## ABSTRACT

For modern Gas sensors, high sensitivity and low power are expected. This paper discusses design, simulation and fabrication of new Micromachined Thermal Converters (MTCs) based on GaAs developed for Gas sensors. Metal oxide gas sensors generally work in high temperature mode that is required for chemical reactions to be performed between molecules of the specified gas and the surface of sensing material. There is a low power consumption required to obtain the operation temperatures in the range of 200 to 500 °C. High thermal isolation of these devices solves consumption problem and can be made by designing of free standing micromechanical hot plates. Mechanical stability and a fast thermal response are especially significant parameters that can not be neglected. These characteristics can be achieved with new concept of GaAs thermal converter.## 1. INTRODUCTION

Standard micro hotplates are based on membranes made of silicon nitride and oxide therefore, the operating temperature is limited to a maximum of about 350°C. The micromachined thermal converters (MTCs) based on GaAs seem to be very attractive for micro hotplates design. In general, MTC integrates GaAs microelectronic devices (high-speed transistors or resistors) and temperature sensors on GaAs thermally isolated micromechanical hotplate. Pt micro-heater, placed at the top of micro hotplate, is designed to warm up sensing surface to operating temperature. Temperature sensors are integrated within hotplate MTC structure to sense the temperature at precisely defined place.

Due to a higher thermal resistance and operation at high temperatures, MTC based on GaAs should be able to perform electro-thermal conversion with higher conversion efficiency than well known Si devices. The most considerable advantages of GaAs, over Si, are some intrinsic material properties such as lower thermal conductivity, high temperature performance, heterostructure quantum effects, etc. The HEMT technology creates conditions for MEMS device development fully compatible not only with signal conditioning and drive circuit but also with the monolithic microwave integrated circuits (MMICs). The most of GaAs based MTC devices were developed to be applied for RF and microwave power sensors and infrared thermal sensors [1]. In this work we demonstrate thermal performance of GaAs based hotplate MTC device designed for gas sensing.

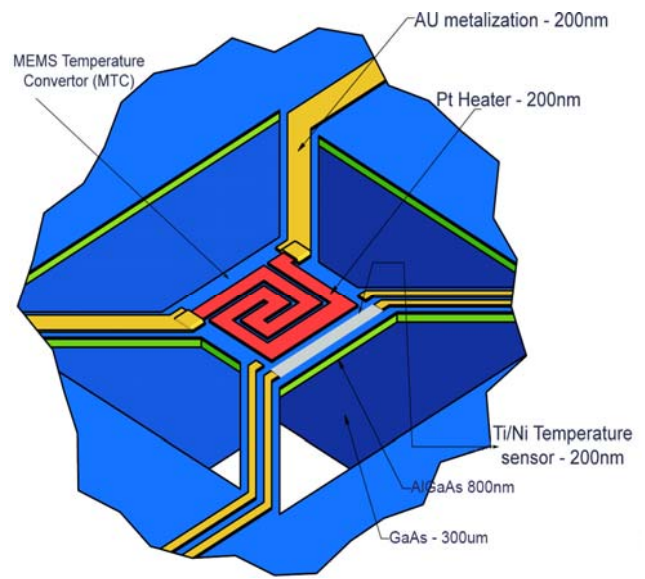

Fig. 1. Model of the MTC suspended hotplate structure. GaAs/AlGaAs hotplate is 2 um thick. SiC barrier layer and gas sensitive NiO layer is not shown.

©EDA Publishing/DTIP 2007                                                                                                    ISBN: 978-2-35500-000-3

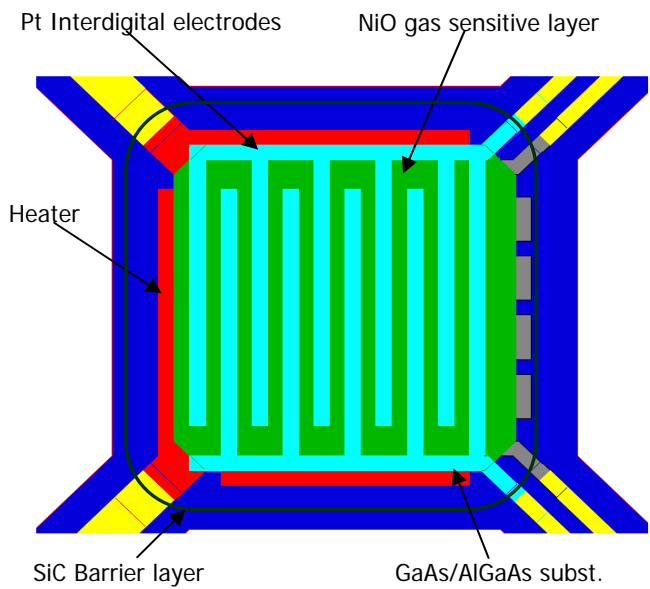

Fig. 2 - Top view mask layout of MTC hotplate

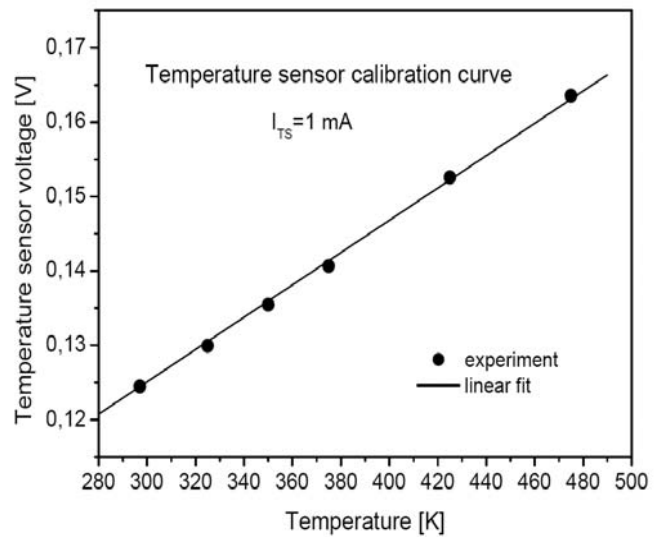

Fig. 3. Measured temperature sensor calibration curve

## 2. 3-D FEM MODEL AND DEVICE FABRICATION

In order to assure excellent thermal isolation of the MTC structure, the hotplate devices are mostly designed as free space standing structures. To increase the thermal resistance values, hotplate has to be designed with the thickness as thin as possible. Additionally, optimization of the MTC structure dimensions, (the aspect ratio between the MTC structure length which increase the thermal resistance and MTC thickness), has to be carried out to find the best trade-off between thermal resistance and acceptable mechanical stress.

Figure 1 shows a schematic view of the hot plate MTC. It consists of Ti/Pt resistor as a heater and Ti/Ni meander-like thin film as a temperature sensor. The both devices are integrated on thermally isolated 2 μm-thick AlGaAs/GaAs island MEMS suspended by the four cross-bridges. The dimensions of GaAs hotplate are 150 μm x 150 μm. For FEM numerical simulation 3-D GaAs model substrate has been designed 10 μm thick and 100 μm wide. Top view mask layout of MTC hotplate is shown in Fig. 2. 500 nm thick SiC barrier layer electrically isolates the 100 nm thick NiO gas sensitive layer and Pt interdigital electrodes.

The MTC hotplate fabrication process begins with the front-side processing technology of the micro-heater and temperature sensor. The process must be combined with surface and bulk micromachining of GaAs and must be fully compatible with the processing technology of integrated microelectronics devices.

The multilayer GaAs/AlGaAs heterostructure active MTC layers are grown by MBE on GaAs substrate. Then, the double-sided aligned photolithography is carried out to define the etching masks on the both sides of the substrate. Highly selective reactive ion etching (RIE) of GaAs from the front side defines the lateral dimension of the hotplate MC structure. Vertical dimension is defined by deep back side RIE through a 300 μm thick GaAs substrate to the AlGaAs etch-stop layer. Consequently the hotplate thickness (vertical dimension) is precisely determined by the depth of MBE grown GaAs layer over AlGaAs etch-stop layer. In the final step AlGaAs etch stop layer is selectively etched. More technological details can be found in [1].

## 3. ELECTRO-THERMAL PERFORMANCE ANALYSIS

The temperature sensitivity of Ti/Ni thin film temperature sensor was investigated in the first stage. I-V characteristic of the temperature sensor at constant current biasing was used to convert the temperature into voltage. Fig. 3 shows the measured voltage response to the temperature at constant current biasing of 1 mA. As expected there is very good linearity in the sensor voltage response observed.

The linear fit performed on the temperature sensor calibration curve in Fig. 3 allows make transfer of the temperature sensor voltage directly to the temperature. Fig. 4 shows measured power to temperature (P-T) conversion characteristic that can be used to evaluate the conversion efficiency of the MTC structure.

As we can see there is some discrepancy from a straight line observed. After fitting the measured data by a quadratic polynomial regression ( $T = 305.23 + 10.297 P + 0.262 P^2$) it is clear that thermal resistance Rth defined

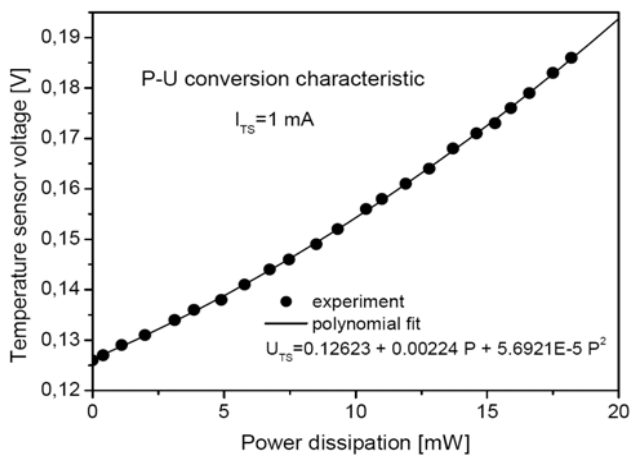

Figure 4. P-T conversion characteristic of MTC device

as ∂T/∂P increases with the power dissipation (temperature increase).

At power dissipation about 20 mW it achieves the value almost 21 K/mW. So, the temperature increase in the sensor active area on the level of 600 K (predicted operating temperature of gas sensor) can be achieved by the power dissipation lower than 20 mW.

Metal oxide gas sensors generally work in high temperature mode that is required for chemical reactions to be performed between molecules of the specified gas and the surface of sensing material. There is a low power consumption required to obtain the operation temperatures in the range of 500 to 700 K. Uniform temperature distribution in the active sensing area is required to ensure equal sensing properties of the whole surface as well.

## 4. DEVICE THERMAL SIMULATION

For an isotropic homogenous material the steady state heat equation can be written [4]:

$$\nabla^2 T \equiv \frac{\partial^2 T}{\partial x^2} + \frac{\partial^2 T}{\partial y^2} + \frac{\partial^2 T}{\partial z^2} = -\frac{1}{k} Q(x, y, z)$$

where $Q$ represents generated internal heat, $k$ denotes the thermal conductivity, $c_p$ its specific heat and $T$ its temperature. The steady state temperature analysis has been performed to determine the temperature distributions and thermal resistance of the MTC device.

For the thermal analysis problem, the essential boundary conditions are prescribed temperatures. The spatial temperature distribution and steady state heat flux were calculated taking into the account the heat transfers to infinity. In the current analysis, according to the application requirement, the fixed thermal boundary is defined for the all side walls of MTC 3-D model. These walls were kept at the room temperature of 300 K while other sides were adiabatic.

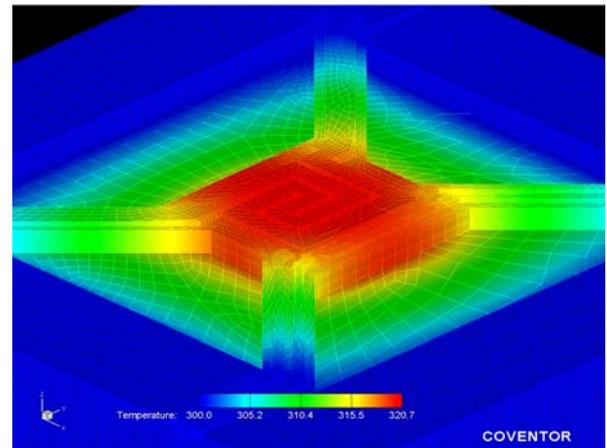

Figure 5. MTC hotplate temperature distribution (Dissipated power in the heater was 1 mW)

3D graph as shown in Figure 5 gives good overall visualization of the temperature distribution in the suspended island structure of the MTC device, which is caused by the power dissipation generated in the thin film resistive Pt heater. The thermal analyses were performed for both vacuum ambient and non-convective gaseous air around the hotplate. The heat losses, due to radiation, were viewed as negligible.

The power to temperature (P-T) conversion characteristics of the MTC device were also investigated by the simulation. High electro-thermal conversion efficiency defined by the extracted thermal resistance value (Rth=17.3 K/mW (CoventorWare simulation) was achieved. This value corresponds to the average value obtained from the experiment (see Figure 4). Transient power characteristics for 1 mW power dissipation are depicted on fig. 6. There are three transients on the fig. 6. Upper is the maximal temperature of the heater and the bottom dependence reflect average temperature of TS.

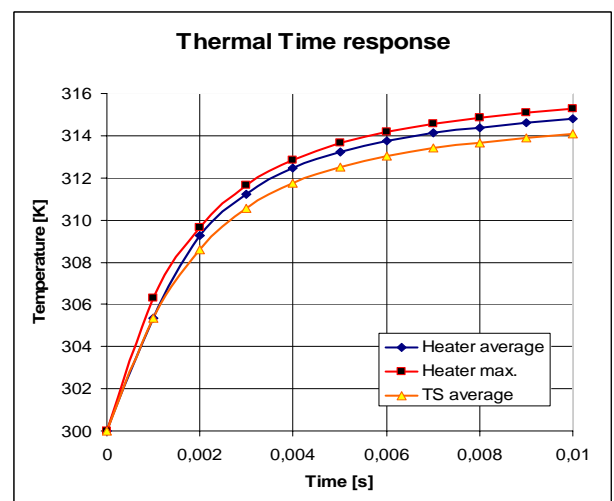

Figure 6. The simulated thermal time response for 1 mW power dissipation in the heater.

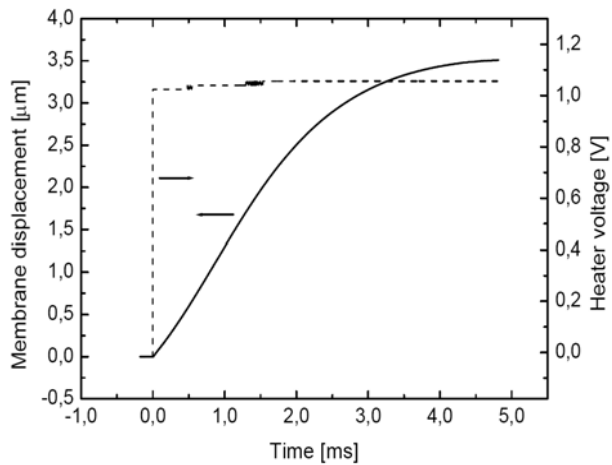

Figure 7. Mechanical time response to the input step-wise heater voltage (temperature increase of 482 K)

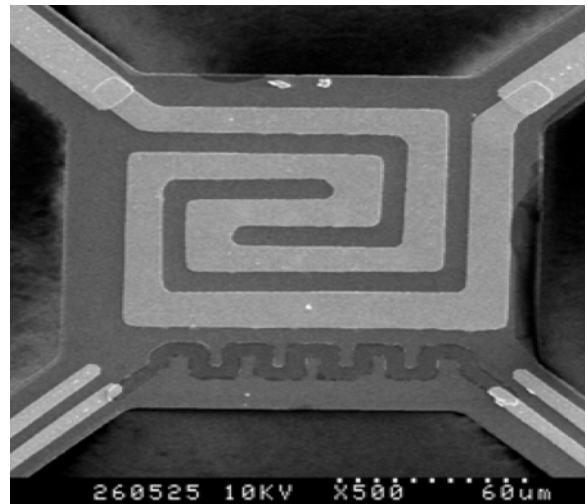

Figure 8. A real view of fabricated MTC device

## 5. DEVICE THERMO-MECHANICAL CHARACTERIZATION

Thermo-mechanical stability and integrity and a fast thermal response belong to very important parameters that can not also be neglected. In order to evaluate the temperature time constant of the MTC device an optical measurement method can be used. It is based on deflection changes measurement of the MTC hotplate which are induced due to different thermo-mechanical properties of the multilayer material system. The non-stationary dynamic process of transient heat flow creates also time dependent mechanical movements. To observe these deformation changes we used Laser Doppler Vibrometer (LDV) optical method. The heterodyne interferometrical system of Polytec OFV-303 vibrometer is capable to detect the vibration amplitudes in nanometer range.

Figure 7 shows the mechanical time response (deflection time dependence) of the suspended hotplate membrane structure obtained by the optical measurement. Extracted thermal time constant value is 1.5 ms and it corresponds to the simulated thermal time constant.

## 6. CONCLUSIONS

The main objective of the presented work was design, modeling and characterization of micromachined GaAs based thermal converter device which is considered to operate with metal oxide gas sensors. The processing technology is fully compatible with the of GaAs MESFET or HEMT devices processing. Subsequently, signal processing electronics can be monolithically integrated with the gas sensors.

Comprehensive electro-thermo-mechanical performance analyses of the MTC hotplate ware performed and exhibit very good mechanical integrity and thermal stability. Due to very high electro-thermal conversion efficiency, defined by the extracted thermal resistance values ($R_{th}$~15-21 K/mW), the power consumption can be kept very low. To obtain the operational temperature of the active part of MTC hotplate in the range of 600-650 K the power consumption was less then 20 mW.

By means of 3-D thermo-mechanical simulation, we optimized MTC hotplate structure to obtain uniform temperature distribution in the active gas sensitive area. Simulated values were compared to experimental values performed by the measurement of real micromachined MTC device. The thermal time constant of the MTC device was also estimated by simulation ($\tau$~1.44 ms) and and compared to experimental measurement ($\tau$~1.5 ms) using LDV method.

The processing technology of described gas sensor based on the MTC device has been in progress. The process flow is now focused to define gas sensitive area based on polycrystalline NiO thin films with a dense fine-grained microstructure.